\documentclass[conference]{IEEEtran}
\IEEEoverridecommandlockouts
\usepackage{cite}
\usepackage{amsmath,amssymb,amsfonts}
\usepackage{algorithmic}
\usepackage{graphicx}
\usepackage{textcomp}
\usepackage{xcolor}
\usepackage[format=plain, belowskip=-13pt]{caption}
\def\BibTeX{{\rm B\kern-.05em{\sc i\kern-.025em b}\kern-.08em
    T\kern-.1667em\lower.7ex\hbox{E}\kern-.125emX}}
\begin{document}

\title{Weighing up the new kid on the block: Impressions of using Vitis for HPC software development}

\author{\IEEEauthorblockN{Nick Brown}
\IEEEauthorblockA{\textit{EPCC, The University of Edinburgh}\\
Bayes Centre, 47 Potterrow, Edinburgh, EH8 9BT, UK \\
n.brown@epcc.ed.ac.uk}
}

\maketitle

\begin{abstract}
The use of reconfigurable computing, and FPGAs in particular, has strong potential in the field of High Performance Computing (HPC). However the traditionally high barrier to entry when it comes to programming this technology has, until now, precluded widespread adoption. To popularise reconfigurable computing with communities such as HPC, Xilinx have recently released the first version of Vitis, a platform aimed at making the programming of FPGAs much more a question of software development rather than hardware design. However a key question is how well this technology fulfils the aim, and whether the tooling is mature enough such that software developers using FPGAs to accelerate their codes is now a more realistic proposition, or whether it simply increases the convenience for existing experts. To examine this question we use the Himeno benchmark as a vehicle for exploring the Vitis platform for building, executing and optimising HPC codes, describing the different steps and potential pitfalls of the technology. The outcome of this exploration is a demonstration that, whilst Vitis is an excellent step forwards and significantly lowers the barrier to entry in developing codes for FPGAs, it is not a silver bullet and an underlying understanding of dataflow style algorithmic design and appreciation of the architecture is still key to obtaining good performance on reconfigurable architectures.
\end{abstract}


\section{Introduction}
Field Programmable Gate Arrays (FPGAs) are configurable integrated circuits that can be programmed to represent, at the electronic gate level, software algorithms. Executing code directly at the gate level, rather than via a general purpose CPU or GPU, provides the opportunity to deliver important performance advantages and potential power efficiencies. Demanding ever more accurate simulations and faster times to solution, scientists and engineers are placing expectations on High Performance Computing (HPC) resources like never before. But whilst scientific ambition expands very rapidly, we are also seeing a stagnation in the reduction of CMOS fabrication size, and with this comes a prediction of an overall deceleration in CPU and GPU raw performance growth. A key question is therefore what role other technologies, and in this paper we focus specifically on FPGAs, can play in accelerating future HPC workloads.

There have been numerous efforts over the years to popularise FPGAs in HPC, however factors such as a reliance on esoteric programming technologies proved to be substantial barriers. In the last few years however, there have been exciting developments at both the hardware (larger, more capable chips) and software (much improved programming environments) levels. This potentially means that the use of FPGAs to accelerate HPC codes is now a more realistic proposition than ever before, and they could be key in obtaining a step change in performance, much like GPUs provided over a decade ago.

In November 2019 Xilinx introduced the Vitis Platform \cite{vitis} which is aimed at making the programming of FPGAs much more a question of software development rather than hardware design. Targeted towards software programmers and users, this platform promises to deliver an environment which is far more familiar to these individuals, and hence lower the barrier to entry in programming FPGAs and their use in accelerating high performing codes. 

Whilst Vitis is in active development, with many updates promised, a key question is whether the current version of this platform, as of April 2020, makes the use of FPGAs in HPC a more realistic proposition. This paper explores that question, and exploits the Himeno Benchmark as a vehicle for examining the programmability properties of the Vitis platform within the context of HPC. The remainder of this paper is organised as follows, in Section \ref{sec:bg} we describe the background to programming FPGAs in more detail, and specifically the Vitis platform, along with a description of the Himeno benchmark. What follows is an exploration of the workflow required when building and running code using the Vitis platform in Section \ref{sec:workflow}. For HPC the ability to optimise code is crucial, which is often relies upon a rich source of insight generated by profiling. Therefore in Section \ref{sec:optimising} we discuss the use of Vitis's profiling tooling to gain critical insights about our port of this benchmark to FPGAs, and explore how such information can be used to optimise performance. Lastly, Section \ref{sec:conclusions} draws some conclusions around the use of this platform for developing HPC codes, and highlights observations that software developers looking to utilise Vitis for FPGA development should be aware of. It is important to note at this stage, that the authors of this paper are independent from Xilinx, and as such are free to provide an honest and fair assessment of Vitis in this text. 

\section{Background}
\label{sec:bg}
Whilst traditionally one would use RTL languages, such as VHDL or Verilog to program FPGAs, there have been numerous efforts to improve the programmability of reconfigurable architectures. High Level Synthesis (HLS), which translates kernel code written in C, C++ or SystemC into the RTL, was a big step in enabling more rapid FPGA development and expanded the user community. Based on HLS, Xilinx proposed their High-Level Productivity Design Methodology \cite{highproduct} which combines HLS with Vivado block design, and it is this that the authors of this paper are most familiar with. Using the methodology, once HLS translates a programmer's code into RTL, a corresponding IP block is then generated, which is imported into the Vivado tool. This tool provides the abstraction of a block design, where the different support functionality including the PCIe bridge, memory controllers, and bus interconnects are all placed and connected. Known as the shell, the programmer then imports their own IP block into this design and must manually connect it to other components and ensure interoperability.

There are two disadvantages to this approach, firstly the programmer must work at the hardware level, understanding interoperability between different block components and on the host side write driver code at a low level, often relying upon OEM specific APIs. The second disadvantage to this methodology is that writing high performance HLS code is a challenge and requires in-depth exploration and experimentation. To do this most effectively the programmer requires a rich ecosystem of tools, including profilers, to understand the performance properties of their code. While there are other programming technologies for FPGAs, such as Intel's Quartus Prime \cite{quartus} for Intel FPGAs and Xilinx's previous SDAccel \cite{sdaccel}, both of which enable host code to be written in OpenCL, these have not reached the level of maturity required for focusing the programming of FPGAs at the software level.

\subsection{Vitis}
The Vitis platform was released in late 2019 and in many ways is the next generation of SDAccel, which itself as a technology has been deprecated. Vitis exposes existing core FPGA development components, such as the tooling provided by HLS and Vivado, as a much more convenient unified development environment. The aim is to enable an approach of not only leveraging reconfigurable computing from the software perspective, but also provide tooling such as profiling and debuggers such that development can be driven in a high level manner. Furthermore, Xilinx have invested heavily in developing a rich set of open-source libraries, documentation, and tutorials that the community are encouraged to contribute towards. These tutorials and examples not only present a walk through of using the platform, but also provide concrete code snippets and examples for a variety of algorithms.

From a software development perspective, hiding the complex aspects of the Vivado tooling, such as partial reconfiguration, enable the programmer to concentrate on their code rather than the esoteric nature of programming FPGAs. Furthermore, the tooling works hand in hand with Xilinx cards, such as the Alveo family, where shells for these cards are supplied and the programmer's generated kernel is automatically integrated by the platform. Not needing to interact at the block design level not only increases productivity, but also enables the programmer to view the architecture as a computational resource rather than hardware system.

For the experiments described in this paper we are using an Alveo U280, which contains over a million LUTs, 32 GB of DDR DRAM memory, 8 GB of High Bandwidth Memory (HBM), and 41 MB of Programmable Logic RAM (PLRAM). The card's runtime provides simple and easy ways of interrogating the status of the FPGA and, whilst it is not strictly speaking part of Vitis, these components interact together seamlessly to automate much of the core management of the card within the Vitis tooling.

\subsection{Himeno benchmark}
The Himeno benchmark \cite{himeno2001himeno} measures the performance of a linear solve of the Poisson equation using a point-Jacobi iterative method. Originally developed to evaluate the performance of CPUs for incompressible fluid analysis codes, the computation required for each grid cell involves 34 single precision floating point operations (13 multiplications and 21 additions or subtractions), and the solver utilises seven data structures; \emph{a}, \emph{b}, and \emph{c} which represent the coefficient matrix with \emph{a} holding four single precision floating point values per grid cell and \emph{b}, and \emph{c} containing three. A further array, \emph{wrk1}, is the source term of the Poisson equation, the \emph{p} array represents pressure, and \emph{bnd} is a control variable for boundaries. Each of \emph{wrk1}, \emph{p}, and \emph{bnd} arrays contain a single, single precision floating point value per grid cell, and there is an additional result array that the calculations are written into. The benchmark reports performance in (single precision) Million Floating Point Operations Per Second (MFLOPs).

It is the mix of computation and data movement that we believe makes the benchmark interesting for this work. Efficient data movement is key to getting good performance on FPGAs \cite{brown2019s}, but also requires more manual control than software developers are commonly used to on CPUs or GPUs. A key question is therefore how well the Vitis platform can guide us in this regard. Some previous studies such as \cite{sato2012evaluating} and \cite{firmansyah2019fpga} have explored the acceleration of this benchmark for FPGAs. For instance \cite{sato2012evaluating} ran on  Maxeler's MAX3 acceleration card (Virtex-6 FPGA) and their Java based programming environment. The authors demonstrated impressive single kernel performance figures for the time of 2700 MFLOPs running entirely from the on-chip BRAM. Differently to their approach, which relied on streaming between the host and device, in this paper we structure the kernel such that all the input data is initially transferred to the card, the kernel executed, and then result data copied back to the host. The reason for this is that it places more strain on the link between the kernel and on-card memory, along with the dataflow aspects of our HLS code, and it is these that we are interested in optimising based upon the insights provided by Vitis.

\section{Building and running with Vitis}
\label{sec:workflow}
In comparison to explicitly using HLS and Vivado block design, as per Xilinx's high-level productivity design methodology, Vitis is driven by the command line with code being built using the \emph{v++} script. Not only is this far closer to what software developers are already familiar with, but there is also some parity with common compiler arguments. This includes \emph{-O} levels which instruct the tooling to perform automatic optimisation during building. Using the \emph{v++} tool one compiles each of their HLS kernels into object files, and these are then linked together, again using \emph{v++}, into a final package, which in this case is a bitstream rather than executable. In addition to promoting familiarity, this also works with standard software development tools, such as make, and is much more streamlined than having to interact with Vivado explicitly.

Building upon the capabilities of SDAccel, Vitis enables programmers to write their host code in OpenCL which is more convenient than previous approaches that often demanded superfluous boilerplate. This also abstracts the programmer from lower level architectural details, enabling the expression of dependencies between kernels and data in a simple and standard manner. Vitis also handles the mapping of data transfers to their corresponding memory spaces, for instance DRAM, HBM, or PLRAM, based upon the context of a transfer and device configuration. As a whole, Vitis significantly reduces the complexity of writing host level code, although one must conform to the OpenCL standard, for instance it is not possible to directly return a scalar value from the kernel, as is the case with direct kernel interaction. Furthermore, unlike Intel's Quartus Prime, in Vitis the device code need not be written in OpenCL, although Vitis does support this, and instead can still follow Xilinx's HLS style.

One of the most welcome features from our perspective is the more convenient way in which emulation is provided by Vitis. The platform provides both software and hardware emulation, and from a compilation perspective the only difference is a single command line argument and then configuration of an environment variable. Moreover, the host and device code can remain unchanged between different emulation modes, although especially for hardware emulation one might wish to reduce the problem size due to the considerable runtime involved. In comparison against using HLS and Vivado block design, via the high-level productivity design methodology, this approach feels much simpler and more convenient. Bearing in mind the sizeable bitstream build time required to run on the actual hardware, it is now much easier to quickly develop code by iterative improvement, relying on emulation to test and validate correctness in the short term, and running on the FPGA less frequently. Whilst the real hardware is often needed for accurate performance measurement, hardware emulation does also provide some estimates about performance too. 

A limitation is that Vitis software emulation is prone to throw cryptic error messages at runtime due to underlying issues in the HLS kernels that have not been identified by the tooling. This is most likely because it does not build the full RTL from the HLS kernels in software emulation mode, but instead performs a short processing phase and then executes in software. We found that one has to compile for hardware emulation mode, which does build the kernel's RTL, to identify many of the potential code level errors. As such, even though it takes slightly longer to build for hardware emulation (for the Himeno benchmark around 7 minutes vs 1 minute), compiling for hardware emulation rather than software emulation is important during development when large code changes have occurred.

Once built, the deployment of a bitstream onto the FPGA is more convenient than traditional approaches, which could involve manually flashing the card and a restart. Vitis shares the approach adopted by Intel's Quartus Prime, where the bitstream is automatically launched on the card via OpenCL calls in the host application. Furthermore there are utilities, such as \emph{xbutil}, which provide card management and metrics.

Generally we found the platform reliable, but there were a small number of instances where Vitis threw unexpected errors, for instance segfaults or licencing errors. We found in all cases that by simply rerunning the command in question the error did not recur.

\section{Optimising performance with Vitis}
\label{sec:optimising}


\begin{figure}
\begin{tabular}{ | c c c | }
\hline
\textbf{Description} & \textbf{MFLOPs} & \textbf{Power (W)}\\  \hline
CPU core (Skylake Xeon) & 3754.49 & 42.30 \\ \hline
Initial FPGA version & 77.82 & 27.20 \\ 
Split out ports & 220.23 & 30.90 \\ 
Memory burst transfers & 301.58 & 30.20 \\ 
Initial 512 bit width & 357.21 &  30.50 \\ 
Bug-fixed 512 bit width & 1452.13 & 31.60 \\
Removed pipeline stalls & 5773.25 & 32.10 \\ 
Increase frequency to 450Mhz & 8658.42 & 33.90 \\\hline

\end{tabular}
\caption{Performance and power usage of Himeno benchmark as optimisations were applied based on Vitis profiling}
\label{fig-kernel-performance}
\end{figure}
Using version 2019.2 of the Vitis platform with an Alveo U280, compiling at optimisation level three for both the host and device code, and GCC version 7.4, we explored the performance of the Himeno benchmark with the middle problem size of x=256, y=128, z=128 over 200 iterations. Figure \ref{fig-kernel-performance} provides an overview of the performance in MFLOPs and power usage in Watts obtained by different configurations of this experiment. The power usage is the entire usage of either the FPGA or CPU, and for context the CPU idles at around 18 Watts. All reported kernel versions are running on the FPGA hardware and timings include both kernel execution and data transfer between the host and device. Host to device data transfer accounts for a tiny percentage of the overall runtime, less than 0.05\%, so we concentrate in this section on optimisation at the kernel level. For comparison the standard, CPU, version of this benchmark was initially run on a single core of Skylake Xeon (single threaded) which delivered a performance of 3754 MFLOPs. Whilst it might seem naive to focus on a single core, we are most interested in the performance optimisations that Vitis can enable at the FPGA kernel level, and as such believe that comparing a single FPGA kernel against a single CPU core is most realistic for this purpose. In contrast to work such as \cite{firmansyah2019fpga} and \cite{sato2012evaluating}, where FPGA versions of this benchmark used either Jacobi solver kernels or a single fat kernel with significant internal parallelism via wide vectorisation, for brevity we limit our discussions around optimisation of, by reducing the overhead within, a single Jacobi kernel with no internal vectorisation. We believe this is reasonable because the tricky part is often minimising overhead within a kernel to obtain good performance, and it is this aspect that we are interested in exploring how well Vitis can assist with by identifying bottlenecks.

\begin{figure}
\centering
\includegraphics[scale=0.35]{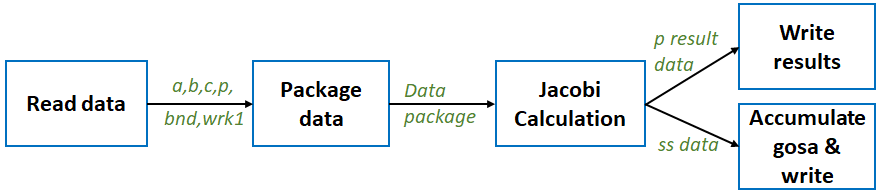}
\caption{Himeno kernel dataflow structure, streaming values between stages, each of which is pipelined internally}
\label{fig:dataflow}
\end{figure}

Our initial FPGA version comprised of a single HLS kernel with five separate dataflow regions running concurrently and connected via HLS streams. Illustrated in Figure \ref{fig:dataflow}, each stage operates on the grid and provides data between stages on a cell by cell basis. The first stage, \emph{read data}, reads grid cell values for each of the six input data structures and via HLS streams of depth 16 these are then passed on a cell by cell basis to \emph{package data} which builds up a data package structure. This structure contains all values needed for calculation on a single grid cell and includes 19 values for \emph{p}, required due to the box stencil. Each data package is then passed to the \emph{Jacobi calculation} stage which performs the calculations involved for the Jacobi iteration for each grid cell as its data package arrives. The result of this for each cell is then passed to the \emph{write results} stage which writes back to memory. The Jacobi calculation stage also passes the \emph{ss} resulting value, used for calculating the residual \emph{gosa}, to a separate stage which accumulates the value for each cell and upon completion of each full iteration writes a single floating point result value to memory. We found that this separation of the physical memory access from data generation or consumption can optimise memory accesses (for instance reducing the number of write or read requests issued), and it also provides a more complex code as a vehicle for exploring the tooling. 

As each stage is running concurrently then effectively there is one very long pipeline, where each dataflow stage is continually passing data to the next stage, and the loops contained within each stage are also running as pipelines internally. The performance of our initial version of this kernel, as described in this section, was \emph{77.82} MFLOPs, around 49 times slower than a single core of the Skylake Xeon CPU. Assuming a clock frequency of 300Mhz and perfect conditions where the pipeline is filled for the entire run and there are no stalls due to factors such as external memory bottlenecks, we estimate that the theoretical peak performance of a single, non-vectorised, Himeno benchmark kernel on the FPGA is around 10000 MFLOPs, which the initial version fell far short of!

Unless otherwise stated, all profiling discussed in this section is based upon runs made on the actual FPGA hardware, and enabling this simply involves a number of additional flags be provided to Vitis during compilation and linking. Upon code termination, CSV files are written by Vitis and these can be loaded up into the Vitis analysis tool for detailed exploration, or even read directly via a text editor. When using this profile to examine the performance of our initial kernel version it was found that the kernel was stalling for over 98\% of the runtime due to external memory accesses. This could be further seen in the application timeline, which is part of the Vitis analysis tool, where \emph{external memory stall} was depicted across the entire execution of the kernel and from zooming in it could be seen that individual accesses to external memory were occupying tiny slivers of space between these stalls.

Vitis analyser also provides advice, known as the guidance pane, for optimising code based upon the results of profiling. This highlighted low data transfer speeds that we were experiencing between the kernel and device memory. Whilst the suite reported that Vitis had automatically selected the use of High Bandwidth Memory (HBM), rather than the slower on-card DRAM, it also highlighted that we were only managing a reading rate of 27 MB/s (0.2\% of bandwidth) and write rate of 50 MB/s (0.4\% of bandwidth). However we also noticed from profiling that, whilst there were six kernel input variables and two output variables, they were all sharing the same single kernel port. As such, we added the \emph{bundle} decorator to the \emph{interface} HLS pragma to create individual kernel ports for each variable. There were two reasons for this, firstly as the HLS kernel can only issue one access per cycle on a port, so with many variables sharing the same port this severely limited the concurrency of reads and writes within our kernel across the input and output variables. Secondly, the HBM on the Alveo U280 is partitioned into thirty two, 256 MB chunks across two banks. The HBM itself contains sixteen memory controllers, where each controller services two of these 256 MB chunks and links to each chunk via an explicit channel. 

Therefore, by adopting separate HLS kernel ports and connecting each to different HBM chunks, we hypothesised that the HBM would be better utilised. This is because Vitis defaults to using the first HBM chunk only, with all accesses going through a single memory controller and channel. This optimisation almost trebled the performance of our initial FPGA version, \emph{split out ports} in Figure \ref{fig-kernel-performance}, and profiling reported that the aggregate read bandwidth had now increased to 429 MB/s. Whilst still nowhere near the theoretical HBM peak bandwidth, by splitting apart and distributing HBM access we significantly increased performance. This is explained in detail in \cite{u280}, but the guidance of Vitis analyser only reported the memory bandwidth and fact that HBM was being used. It could be said that Vitis was somewhat misleading here, as the guidance pane reported \emph{healthy} against the fact that HBM was being used, with no indication that a simple configuration change could provide much greater bandwidth. This is especially invidious bearing in mind a software developer, who likely has no experience in interacting with memory at this low level, will be prone to accept the guidance of the tooling which, in this situation, would result in performance degradation. On the plus side, only a trivial configuration change was needed to split across the HBM memory, and no changes were required to the code. This demonstrates the benefit of the abstraction provided by Vitis and the ability to easily experiment with different tuning options.

At this point profiling still reported that a large proportion of time was being spent in HBM memory access, and the guidance pane advised that we should do two things; increase the number and size of burst transfers, and increase the data width of our kernel ports from 32 bits (single precision floating point) to 512 bits. To improve the burst transfers we added appropriate qualifiers, for instance \emph{num\_read\_outstanding} and \emph{max\_read\_burst\_length} for reading, to the AXI4 interface pragmas in HLS and all interactions with HBM were driven via internal PLRAM memory, unpacking or packing this as required by the kernel. This optimisation increased performance to 301 MFLOPs, \emph{memory burst transfers} in Figure \ref{fig-kernel-performance} and Vitis profiling reported that the aggregate read bandwidth had increased to 1 GB/s, but still with a substantial number of memory stalls, around 70\% of the overall kernel runtime.

Following the guidance of Vitis analyser we next refactored the HLS code to increase the width of kernel ports connected to the HBM from 32 to 512 bits. This was achieved by packing 16 single precision floating point values into a C structure and applying HLS's \emph{DATA\_PACK} directive. Whilst it added somewhat to the code complexity, having to unpack these wide structures and consider edge-cases at the boundaries where data ran beyond the area of interest, this not only followed the guidance of Vitis but also more general best practice \cite{brown2019s} to fully utilise the memory controllers. We were initially disappointed with the performance improvement afforded by this optimisation, \emph{initial 512 bit width} in Figure \ref{fig-kernel-performance}, delivering only 357 MFLOPs. 

Using Vitis to understand why, we found that we had made a simple coding error which, from the profiler, was immediately obvious. A burst length of 1.024 KB was reported for the \emph{p}, \emph{bnd}, and \emph{wrk1} fields, with the guidance providing a healthy indication, but for the \emph{a}, \emph{b}, and \emph{c} fields the burst size was only 0.004KB and guidance flagged this as problematic. We found that it was simply a case of accidentally omitting the \emph{DATA\_PACK} directive for these three fields in the HLS code, and the addition of this pragma considerably improved the performance to 1452 MFLOPs. At this point profiling reported an aggregate read bandwidth of 61 GB/s, with all individual kernel ports reporting a bandwidth utilisation of around 90\%. Furthermore, profiling data reported that memory stalls now accounted for only 0.06\% of the overall runtime, which was confirmed by examining the Vitis timeline trace.

We were now confident that, by using the insights provided by the Vitis platform, the overheads associated with memory stalls to and from the HBM had been addressed. However, we were still achieving less than half the performance of a single Skylake Xeon CPU core. Whilst the Vitis profiling tool had been very useful, it was unable to provide more insight. Namely, whilst the profiler can provide detailed information external to the kernel HLS IP, it is more limited inside the IP and, even though Vitis reported less than 0.001\% of runtime lost due to intra-kernel dataflow stalls, a question was how well the different dataflow regions inter-operated and for how much time our pipeline was fully filled. A concern was that different dataflow regions of the HLS code could be poorly load balanced, causing an excessive number of stalls internally, but unfortunately whilst Vitis profiling provides some information around intra-kernel stalls, it does not break this down to the granularity required to understand where exactly stalls are occurring, and how well occupied are the pipelined loops.


Until this point we had relied on profiling based upon runs on the FPGA, rather than hardware emulation. The reason is that in our experience, emulation profiling is overly optimistic about memory accesses, reporting 100\% utilisation on all kernel ports for instance. However, hardware emulation provides additional information, namely the ability to generate a waveform which displays in-depth details at the system, kernel, and function level. This additional profiling tool includes details such as data transfers between kernel, global memory access, and data flow through inter-kernel pipes. The idea of this is to provide insight into performance bottlenecks at the individual function calls level, however it does not report information about intra-kernel pipes.

We therefore realised that we would obtain far more insight about performance of our code by splitting apart the dataflow functions of our single, monolithic, kernel into separate HLS kernels connected via inter-kernel HLS streams. The hypothesis was that this refactoring effort would result in a structure more easily measured by the profiling tool, with increased information on a kernel by kernel basis, and thus more effectively help us quantify and understand additional kernel stalls. This separation added slightly to the code complexity, for instance our streams were now AXIS kernel ports and of type \emph{ap\_axiu}, requiring some packing and unpacking of data. However, it was very convenient to connect the streams of kernels together in Vitis using the \emph{stream\_connect} linker option via a configuration file. Whilst the Vitis documentation is generally good, it wasn't entirely clear that the order of this configuration matters, i.e. is \emph{producer:consumer}.

Separating the dataflow regions into individual kernels slightly impacted the MFLOPs performance, but we found provided substantially more insight when profiling. Vitis analyser now reported detailed statistics around inter-kernel pipe stalls based on runs carried out on the FPGA and from this data it could be seen that some of the kernels were stalling for up to 25\% of their runtime due to stream stalls, and from looking at the streams themselves the stall rate ranged from 19\% to 60\%. This indicated that there were some crucial inefficiencies, and bearing in mind the good utilisation on HBM, we felt that this was most likely driven by load imbalances in different stages of our implementation. From this we surmise that Vitis profiling is likely much better suited to monitoring at the shell level (e.g. the utilisation of external kernel ports, such as the AXI4 connections to HBM and inter-kernel AXIS streams) rather than profiling within each individual HLS kernel. This same picture could be seen by examining the waveform profile generated from hardware emulation, and the viewer reported that much of the stalling was occurring in the streams of the first stage which connects the reading of data (which was not stalling) to the packaging of this data in the second stage.

Based upon further investigation via the Vivado HLS Eclipse based IDE, we found that HLS was imposing an initiation interval (II) of 4 for the consumption of array \emph{a} data and 3 for arrays \emph{b} and \emph{c}. This corresponds directly to the number of values needed per-grid cell and the reason was that there was a conflict on the single 32 bit wide stream we were using to connect the stages. The consequence was that a downstream stage could only read one value per cycle but in-fact needed up to four per individual grid cell, effectively stalling and only processing a grid cell every four cycles. To resolve this for kernel ports \emph{a}, \emph{b}, and \emph{c} which are connected to the HBM, instead of using a packed structure of sixteen single precision floating point numbers we replaced this with a \emph{ap\_uint} of width 512 bits. Applying the \emph{range} operator, this 512 bits wide \emph{ap\_uint} was then unpacked into chunks of type \emph{ap\_uint} 128 bits wide (4 single precision floats) for \emph{a} and 96 bits wide (3 single precision floats) for \emph{b} and \emph{c}. It was now these \emph{ap\_uint\textless128\textgreater} and \emph{ap\_uint\textless96\textgreater} chunks that were sent between stages via HLS streams, one per cycle, and as such we no longer encountered such a stall between stages or within pipelined loops. This very significantly improved performance on the FPGA, to 5773 MFLOPs which out performs the single Skylake CPU core. It is important to note that this performance figure is the one achieved by our single, monolithic, HLS kernel rather than the separated kernel. This is because it is the monolithic kernel that delivers optimal performance, and as such we found it worked best to exploit the separated version to deliver insights via profiling and then apply the resulting optimisations back into this single monolithic kernel to obtain optimal performance. Whilst this is a somewhat obvious optimisation from the HLS perspective, it was the profiling that gave us confidence that the kernel was no longer stalling on external memory accesses but instead issues within the kernel accounted for the predominant overhead. Based on this information, and using our existing knowledge of HLS, we were able to identify these and address them.

During compilation Vitis reports an estimated maximum clock frequency for each kernel, and this was originally around 330 Mhz for our code. We ran our design at 300 Mhz, which is the default clock speed, but when refactoring the HLS code to reduce the stalls, we discovered that this timing limitation occurred within the stage that calculates \emph{gosa}, the relative residual. In this stage contributions from each grid cell are accumulated, and whilst we had split this out into a completely partitioned temporary array and unrolled the loop, this was based on a factor of 11. The factor was driven by a latency of 11 clock cycles for the \emph{fadd} operation, but by increasing this factor to 20 we gained additional timing slack, at the cost of increased resource usage. Subsequently, when compiling using Vitis the platform reported an estimated maximum kernel clock frequency of 501 Mhz. We therefore increased the clock frequency to 450 Mhz, which matches the frequency of the HBM memory controllers and ensures that we are well within the maximum frequency to meet timing. The result was a performance of 8658 MFLOPs, which is a further noteworthy performance improvement.

Whilst the focus of this section is to use the benchmark as a vehicle for exploring the profiling capabilities of Vitis, it should be noted that the final, 8658 MFLOPs kernel utilises 25889 LUTs, 29406 FFs, 67 BRAM and 25 URAM blocks, and 81 DSP slices. This represents at most 2\% of the Alveo U280 resources, and even when considering the additional requirements of the shell, there are a large number of available resources left to further parallelise this code by running multiple kernels and/or applying internal vectorisation.

\section{Conclusions and observations}
\label{sec:conclusions}
In this paper we have explored the use of Xilinx's new Vitis platform for building, executing, and profiling HPC codes, driving the discussion via the Himeno benchmark. Whilst there is plenty to further explore in Vitis, for instance further work examining debugging capabilities and reusability of open-source kernels developed by Xilinx, we can still form a number of conclusions. From a software development perspective, Vitis is a considerable improvement over more traditional FPGA programming approaches, but there are still some limitations. Driving software development via the command line and Vitis alone will likely result in correct code, but to obtain good performance more knowledge and insight is still required and this can be leveraged efficiently using the platform. 

The paramount observation is that to gain good performance the programmer must still leverage the insights provided by the HLS tooling, specifically the detailed logs generated and schedule explorer in the Eclipse based IDE. We feel this is important to stress because performance on FPGAs is so closely tied to developing appropriate dataflow algorithms, and with Vitis code can be written using any IDE and then compiled. However to obtain good performance there is no substitute for using Xilinx's HLS IDE, synthesising often and examining the results via the schedule explorer to improve ones HLS algorithm. Whilst it is possible to load the summary of HLS compilation into Vitis analyser, in our opinion this is not prominent enough, and a danger is that programmers do not realise the impact of their suboptimal HLS code.

More generally, Xilinx should continue to develop the profiling capabilities of Vitis, focusing more at the intra-kernel level. HPC programmers are used to being able to track the proportion of runtime, often on a line by line basis, accounted for by different parts of their code. Even being able to explore statistics around the percentage time each pipelined loop is fully filled, partially filled, and stalled would provide important insights around performance bottlenecks. 

It is our opinion that Vitis is a significant step in the right direction, and as it currently stands is accessible for HPC software programmers to write correct, but not necessarily high performance, codes running on an FPGA. In order to gain high performance on FPGAs a programmer at this time must still have a deep understanding of how to write dataflow style algorithms and an appreciation of the underlying architectural details, but can use the Vitis platform to direct their efforts, rather than having to address all the low level and tricky details manually. This makes them more a director, rather than labourer, of FPGA programming and as Vitis continues to evolve it has the promise to ultimately render widely accessible the acceleration of HPC codes using reconfigurable computing.

\bibliographystyle{./bibliography/IEEEtran}
\bibliography{./bibliography/IEEEabrv,./bibliography/IEEEexample}

\end{document}